\documentclass[a4paper,aps,pra,twocolumn,showpacs,preprintnumbers,amsmath,amssymb]{revtex4}
\usepackage{bbm}
\usepackage{amsmath}
\usepackage{verbatim}
\usepackage{graphicx}

\def\RR{\ensuremath{\mathbbm{R}}}
\def\CC{\ensuremath{\mathbbm{C}}}
\def\id{\ensuremath{\mathbbm{1}}}

\def\cN{{\cal N}}
\def\cP{{\cal P}}
\def\cS{{\cal S}}

\def\tr{\mathrm{tr}}
\def\ket#1{\left| #1\right>}
\def\bra#1{\left< #1\right|}
\renewcommand{\Re}{\mathrm{Re}}

\newcommand{\Eqref}[1]{Eq.~(\ref{#1})}

\newcommand{\Sref}[1]{Sec.~\ref{#1}}
\newcommand{\Aref}[1]{Appendix \ref{#1}}
\newcommand{\Fref}[1]{Fig.~\ref{#1}}
\def\qed{\rule{1ex}{1ex}}
\def\Ubs{U_\mathrm{bs}}
\newcommand{\acosh}{\mathrm{acosh}}
\newcommand{\diag}{\mathrm{diag}}

\def\sev{\lambda_{\mathrm{min}}} 
\def\ssv{\sigma_{\mathrm{min}}} 
\def\gamt{\tilde\gamma}
\def\delt{\Delta t}
\def\sqm{Q}
\def\entm{E}

\def\be{\begin{equation}}
\def\ee{\end{equation}}
\def\bi{\begin{itemize}}
\def\ei{\end{itemize}}
\def\bea{\begin{eqnarray}}
\def\eea{\end{eqnarray}}
\def\bma{\begin{mathletters}}
\def\ema{\end{mathletters}}

\def\C{\hbox{$\mit I$\kern-.7em$\mit C$}}

\newcommand{\one}{\mbox{$1 \hspace{-1.0mm}  {\bf l}$}}
\def\one{\id}
\newcommand{\proj}[1]{\ket{#1}\bra{#1}}

\newcommand{\citfn}[2]{
\newcounter{#1}\footnote{#2}\setcounter{#1}{\value{footnote}}}
\newcommand{\cfn}[1]{\footnotemark[\value{#1}]} 
\begin{document}
\title{Entanglement generation and
Hamiltonian simulation in Continuous-Variable Systems}

\author{Barbara Kraus, Klemens Hammerer, G{\'e}za Giedke, and
J. Ignacio Cirac}

\affiliation{Max-Planck--Institut f\"ur Quantenoptik,
Hans-Kopfermann-Strasse, D-85748 Garching, Germany}

\begin{abstract}
Several recent experiments have demonstrated the promise of atomic
ensembles for quantum teleportation and quantum memory. In these
cases the collective internal state of the atoms is well described
by continuous variables $X_1, P_1$ and the interaction with the
optical field ($X_2, P_2$) by a quadratic Hamiltonian $X_1X_2$. We
show how this interaction can be used optimally to create
entanglement and squeezing. We derive conditions for the efficient
simulation of quadratic Hamiltonians and the engineering of all
Gaussian operations and states.
\end{abstract}


\maketitle

\section{Introduction}
After the first experiments \cite{FuSBFKP98} on quantum
teleportation using two--mode squeezed states
\cite{Vaid94,BrKi98}, as well as those
\cite{HaSSP99,JuKP00,KuMB00} dealing with entanglement in atomic
ensembles \cite{KuPo00,DuCZP00}, a significant amount of work has
been devoted to develop a quantum information theory of continuous
variable systems. So far, most of the theoretical work has focused
on the entanglement properties of the quantum states involved in
all these experiments, the so--called {\em Gaussian states}. Some
examples of the achievements in this field are the following. The
problem of qualifying entanglement has been solved in the general
bipartite setting \cite{DuGCZ99,Simo99,WeWo00,GiKLC01} and in the three
mode case \cite{GiKLC01b}. The distillation
problem has also been answered in the general case
\cite{GiDZC01}, as well as in the case in which the class of
allowed operations is restricted to those that conserve the
Gaussian form \cite{GiCi02,EiSP02,Fiur02}. In contrast to all
this theoretical work on (the static) entanglement properties of
Gaussian states, very few results \cite{ThMW02,BoMo02,dLMo02,BeSa02}
have been obtained on the dynamics of entanglement on these systems,
i.e., on how to use the interactions provided by the physical set--ups
in order to entangle the systems in the most efficient way.  This
paper provides a rather complete theory of
the dynamics of entanglement in these experimental settings.

The dynamics of entanglement has been recently analyzed in systems
of two or more qubits
\cite{DuVCLP00,KrCi00,ZaZF00,DoNBT01,WoRJB01,NiBDCD01,BeHLS02,LeHL02}.
In that case one distinguishes between two scenarios. In the first
one \cite{DuVCLP00,ZaZF00}, the interaction between the qubits is
described by a Hamiltonian $H$. The goal is to determine the
sequence of local gates for which the increase of entanglement
after some small (infinitesimal) time is maximal for a given
initial state. In the second one \cite{KrCi00,DoNBT01,LeHL02}, the
interaction is given in terms of a non--local gate, which can be
applied only once. Apart from its fundamental interest, these
studies give some practical ways of creating entanglement in the
most efficient way and may become relevant in several experimental
situations. Another interesting and related problem is the one of
Hamiltonian and gate simulation
\cite{ViCi01,BeCLLLPV01,ViHC01,HaVC02}. Here, one assumes that the
two qubits interact via some given Hamiltonian $H$ and the goal is
to determine a sequence of local instantaneous gates in order to
obtain in minimal time either a complete time-evolution generated by
some other Hamiltonian [Hamiltonian simulation] or some desired
unitary gate (gate engineering).

In the present paper we analyze all these problems for two--mode
pure Gaussian states and interaction Hamiltonians which preserve
the Gaussian character. We also study the generation of squeezing,
since although it has no counterpart in the qubit case, it is a
valuable resource in present experiments \cite{WoEP02}. Given the
fact that we touch on several different topics and therefore
develop different mathematical tools, we have decided to write a
section which explains in detail the different problems we
consider and the corresponding results. In the following sections
we give detailed derivations of these results.

We stress the fact that the problems studied here are all motivated by
the experimental situation in which light
gets entangled with an atomic ensemble via a Kerr--like
interaction \cite{KuBM98,Molm99,Polz99,DuCZP00}. We expect that
the techniques developed in this paper can be easily extended to
address other related problems, like the one of entangling two
atomic ensembles using light.

The paper is organized as follows: The section \Sref{SecOver}
should be considered as a survey of the results presented in the
paper. In Section \Sref{SecSim} we show which Hamiltonians can be
simulated using a given interaction and how to do so optimally. We
also show that, in fact, any general Gaussian operation can be
generated in the considered set--up. In Section \Sref{SecRate} we
determine the optimal rate of entanglement generation as well as
of squeezing generation for arbitrary input states. Finally, in
Section \Sref{SecFinTim} we give an optimal entanglement
generation scheme for finite times, starting out from a product
(unsqueezed) state.

\section{Overview}\label{SecOver}

This section gives an overview of the content of this paper and it
is composed of three subsections. In the first one, we explain the
physical set-up that we are going to analyze. In the second one we
collect the main definitions used thereafter. In the third section we
give the main results of the paper without proving them. For the
detailed derivations we refer the reader to the following sections.

\subsection{Setup}

We consider a continuous variable system composed of two one--mode
systems coupled via some interaction Hamiltonian. The goal is to
analyze which kind of evolutions we can achieve with such an
interaction if certain instantaneous local operations can be
applied at will. In particular, we study optimal methods of
creating or increasing the entanglement shared by the two modes.

The interaction Hamiltonian has the general form
\begin{equation} \label{generalHint}
H=aX_{1}X_{2}+bP_{1}P_{2}+cP_{1}X_{2}+dX_{1}P_{2}
\end{equation}
where $a,b,c$ and $d$ are real parameters, and $X_{1,2}$ and
$P_{1,2}$ are canonical operators for the first and second mode,
respectively \citfn{fnote0}{Note that the Hamiltonians that we are considering here
are not semi-bounded. However, this is not a problem since we are
always considering initial states and times for which the real
Hamiltonian can be locally approximated by these Hamiltonians.}.
We use dimensionless units throughout the paper.
We assume that local operations, generated by the
Hamiltonians
\begin{equation}\label{Hloc}
H_{\mathrm{loc},i}= g(X_{i}^{2}+P_{i}^{2}),
\end{equation}
can be applied instantaneously, where $g$ is a real number
that can be tuned at will \cfn{fnote0}. These operations can neither
change the entanglement nor the squeezing present in the
state. Lastly, we assume that the initial state is pure and Gaussian.

Our choice of the Hamiltonian interaction as well as the instantaneous local
operations is motivated by current experiments
with atomic ensembles \cite{ScJSP02,JuKP00,KuMB00,HaSSP99}.
In particular, to those set-ups in which an atomic ensemble interacts
with two modes of the electromagnetic field
\citfn{fnote1}{This is the case if the atomic ensemble
is embedded in a ring cavity, for example. Note that for optically
thick samples, the description may also be valid in free
space.} with different polarizations
\cite{KuBM98,KuPo00,DuSCZ00,DuCZ02}. If the atoms are
sufficiently polarized along some given direction (say $x$) we can
replace the total angular momentum operators describing the
internal state of the atoms by canonical operators. That is (if the
involved levels have spin $\pm1/2$), $S_y\to X_1/\sqrt{N/2}$, $S_z\to
P_1/\sqrt{N/2}$, $S_x\to N/2$,
with $[X_1,P_1]\simeq i$ ($\hbar=1$), and where $N$ is the number
of atoms. This approximation is valid as long as $|\left<S_x\right>-N/2| \le
o(\sqrt{N})$ for all times \cite{ArCGT72}. Similarly, if the electromagnetic
field is sufficiently polarized along some direction, we can
substitute the Stokes operators by canonical ones, $X_2$ and
$P_2$ \cite{ScBTRBL02}.

For some atomic structures and off-resonant interactions, the
Hamiltonian describing the interaction between the atomic ensemble
and the light can be written as \cite{KuBM98}
\begin{equation}\label{Hint}
H_0= a X_{1}X_{2},
\end{equation}
which is a particular case of \Eqref{generalHint}; in the following we
will put the coupling constant $a=1$ when referring to $H_0$. In the same
scenario, simple and fast local operations can be performed on the
atoms and the electromagnetic field. For example, a magnetic field
or a polarizer gives rise to the local Hamiltonians \Eqref{Hloc}.
Since the interaction between atoms and light is typically weak,
with moderate magnetic fields the operations generated locally can
be regarded as instantaneous. On the other hand, if the atoms and
the light are completely polarized, the corresponding state in
terms of our continuous variable description is the tensor product
of two vacuum states, in particular it is a pure Gaussian state.

We emphasize that even though we have motivated our choices with
some particular physical set-up, our description is applicable to
other physical situations and our results apply to the 
general interaction Hamiltonian \Eqref{generalHint}. In that case,
we make no more references to the physical nature of our systems.
However, in some cases we particularize our results to the
considered physical situation described above.

Now we consider the following general \emph{strategy} for state
or gate engineering which can be realized using the tools described
above. Starting with a pure initial state, described by the density
operator $\rho(0)$, we perform fast local operations $V_0\otimes
W_0$ on the state and we then let $H$ act on it for a time $t_1$.
Then we perform again local rotations, $V_1\otimes W_1$ followed
by the non--local interaction generated by $H$ for a time $t_2$
and so on until $\sum_kt_k=t$. This yields to the total time-evolution
operator
\begin{equation}
\mathcal{U}(t)=[V_{n}\otimes W_{n}]U(t_n)\cdots
U(t_2)[V_{1}\otimes
W_{1}]U(t_1)[V_{0}\otimes W_{0}], \label{Simprot0}%
\end{equation}
so that $\rho(t) = \mathcal{U}(t) \rho(0) \mathcal{U}(t)^\dagger$.
Here $U(t)=e^{-iH t}$.

First, we want to analyze which $\mathcal{U}$ are achievable with
this strategy. Second, for a given $\rho(0)$ we look for the best
choice of $n$, $\{t_1,\dots,t_n\}$, and the local operations
$\{V_1\otimes W_1 ,\dots,V_n\otimes W_n\}$ in order to maximize
the created entanglement/squeezing. We consider two different
regimes. First, we choose $\sum_k t_k=\delta t \ll \tau(H)$ (the
characteristic time of the interaction) so
that we can expand all the $U$ as well as $\mathcal{U}(t)$ in
lowest order in $t_k$. Second, we choose $t_k$ finite. In the
following we refer to those two regimes as infinitesimal and finite
respectively.

\subsection{Some definitions}

Since all the Hamiltonians we are considering are at most
quadratic in $X$ and $P$, an initial Gaussian state will be
Gaussian at all times. This means that we can fully describe it by
the first and second moments of $R_k$, with $\vec R =
(X_1,P_1,X_2,P_2)^T$, i.e. the expectation values $d_k=\tr(\rho
R_k)$, also called displacements of $\rho$ and the variances $\tr[\rho
(R_k-d_k) (R_l-d_l)]$. The 
latter are collected in the \emph{correlation matrix}
(CM) of the state $\rho$, the real, symmetric, positive matrix
$\gamma$  defined by
\begin{equation}\label{CM1}
\gamma_{kl} =  2 \Re\{\tr[\rho (R_k-d_k) (R_l-d_l)]\}.
\end{equation}
In our description, the displacements are of no importance:
they have no influence on the entanglement and squeezing properties of
the states and can be brought to zero by local displacement
operations, which can be easily implemented in our physical
set-up. Therefore we take $d_k=0$ in this paper.

We often write the correlation matrix in the block form
\begin{equation}\label{gamma}
 \gamma = \left( \begin{array}{cc} A&C\\ C^T&B
\end{array} \right).
\end{equation}
with $2\times2$ matrices $A,B,C$, where $A$ refers to the first
system and $B$ to second system. The matrix $C$ describes
the correlations between both systems and vanishes for product
states.

All the states and operations we consider here are pure. Therefore,
and since we look at two-mode states only, we
can always write their CM in the form \cite{SiMD94}
\begin{equation}\label{gamma2}
 \gamma = (S_1\oplus S_2)\left( \begin{array}{cc} \cosh(r) \one& \sinh(r)\sigma_z\\
 \sinh(r) \sigma_z&\cosh(r)\one
\end{array} \right)(S_1^T\oplus S_2^T),
\end{equation}
which we refer to as the \emph{pure state standard form} of
$\gamma$. Here, $S_{1,2}$ are local symplectic matrices, $r\geq0$,
and $\sigma_z$ is the Pauli matrix $\diag(1,-1)$. The parameter
$r$ contains all information about the entanglement of the state,
whereas $S_1$ and $S_2$ contain information about local squeezing.
Given a CM $\gamma$, one can readily find its pure state standard
form \citfn{fnStdForm}{We have $S_k=O_kD_kO_k'$, where $O,O'$ are
rotations and $D_k=\diag(e^{r_k},e^{-r_k})$. The six matrices are
determined as follows: $O_{1(2)}$ diagonalize $A (B)$. Then the
two eigenvalues $\alpha_1,\alpha_2$ of $A$ determine $e^{r_1}$ as
$(\alpha_2/\alpha_1)^{1/4}$ (similarly for $e^{r_2}$) and the $O'$
are the rotations that realize the singular value decomposition of
$D_1^{-1}O_1^TCO_2D_2^{-1}$. The two-mode squeezing parameter $r$
is given by $\cosh r=\sqrt{|A|}$, while the squeezing parameters
$r_1,r_2$ of $S_k$ can then be calculated by the trace of $A$ and
$B$, resp.: $\cosh r_k = (\tr A)/(2\cosh r)$.}.

Concerning the bilinear interaction Hamiltonians, it is convenient
to rewrite the Hamiltonian of \Eqref{generalHint} as follows
\begin{equation}\label{Kform}
H=(X_{1},P_{1})K { X_{2}\choose P_{2}}
\quad\text{where}\quad K=\left(
\begin{array}
[c]{cc}%
a & d\\
c & b
\end{array}
\right).
\end{equation}
We denote by $s_{1}=\sigma_{1}$,
$s_{2}=\mathrm{sign}[\mathrm{\det}(K)]\sigma_{2}$
\citfn{notesign}{The sign function is defined as $\mathrm{sign}
(x)=\pm1$ if $x\gtrless0$ and $\mathrm{sign}(x)=0$ if $x=0$.} with
$\sigma_{1}\geq\sigma_{2}\geq0$ the singular values of $K$. We
refer to the $s_k$  as the \emph{restricted singular values} of
$K$. Note that, local rotations can always bring any $H$ to the
diagonal form $s_1X_1X_2+s_2P_1P_2$.

\subsection{Results}

We state here the main results of this paper. To give a clear
picture of them we do not use more mathematically tools and
definitions than necessary.

First we characterize the interactions which we are able to
generate within the setting described by \Eqref{Simprot0}. In the
infinitesimal regime the problem is usually called Hamiltonian
simulation, whereas for $t$ finite it is usually called gate
simulation. Then we use these results to find the optimal strategy
to generate entanglement/squeezing both in the infinitesimal and
finite regime.

\subsubsection{Hamiltonian Simulation}

Given two Hamiltonians $H$ and $H'$ of the form
(\ref{generalHint}) we want to see the conditions under which $H$
can simulate $H'$. That is, for a given sufficiently small $t'$ we
want to find out if it is possible to have
\begin{equation}
\label{Simprot} e^{-iH't'}=[V_{n}\otimes W_{n}]e^{-iHt_{n}}\cdots
e^{-iHt_{2}}[V_{1}\otimes W_{1}]e^{-iHt_{1}}[V_{0}\otimes W_{0}].
\end{equation}
with $t_k$ small as well. If it is possible to choose $t\equiv
\sum_k t_k=t'$ we say that $H$ can simulate $H^\prime$
\emph{efficiently}.

Defining the matrices $K$ and $K'$ as in \Eqref{Kform}, as well as
their respective restricted singular values $s_{1,2}$ and $s_{1,2}'$, we find the
following results: (i) The Hamiltonian $H$ can efficiently simulate
$H^{\prime}$ if and only if
\begin{equation}
s_{1}+s_{2}\geq s_{1}^{\prime}+s_{2}^{\prime}\quad\text{and\quad}s_{1}%
-s_{2}\geq s_{1}^{\prime}-s_{2}^{\prime},
\end{equation}
(ii) If it is not possible to simulate $H^\prime$ efficiently
with $H$, then the minimal time needed to simulate the evolution
corresponding to $H^\prime$ for the time $t^\prime$ is
$t_{\min}:=\min_t\{t:(s_{1}+s_{2})t\geq
(s_{1}^{\prime}+s_{2}^{\prime})t^\prime,(s_{1}-s_{2})t\geq(s_{1}^{\prime}%
-s_{2}^{\prime})t^\prime\}$.\\
Thus except for the cases $s_1=\pm s_2$ every Hamiltonian of
the form (\ref{generalHint}) can simulate all other Hamiltonians of
that form (including the $s_1'=\pm s_2'$ case).
In particular, with the Hamiltonian $H_0$ describing the
atom-light interaction one can simulate every bilinear Hamiltonian
(\ref{generalHint}) and can do so efficiently as long as
$|s_1'|+|s_2'|\leq 1$.  In this case, the interaction existing in the
physical setup can be considered universal.

\subsubsection{Gate simulation and state generation}

We show that starting from the Hamiltonians $H$ and
$H_{\mathrm{loc},i}$ of Eqs. (\ref{generalHint},\ref{Hloc}) it is possible to
generate any desired unitary evolution of the form $\mathcal{U}=e^{-i\tilde
H}$, where $\tilde H$ is an arbitrary self--adjoint operator
quadratic in $\{X_1,P_1,X_2,P_2\}$, if and only if $|s_1|\ne
|s_2|$. In particular, the Hamiltonian $H_0$ allows to generate
all unitary linear operations, and therefore to generate arbitrary
Gaussian states out of any pure Gaussian state. This shows that
$H,H_{\mathrm{loc},i}$ generate a set of universal linear gates for continuous
variables smaller than the one given in Ref. \cite{LlBr99}.

Let us analyze some important applications of these results in the
case of atomic ensembles interacting with light. They imply that with
current experiments with atomic ensembles one can generate all
unitary linear operations, as well as arbitrary Gaussian states. In
particular, one can generate local squeezing
operators for which $\tilde H=X_1^2-P_1^2$ [which are not included
among the Hamiltonians of the form (\ref{generalHint}) and therefore
cannot be simulated infinitesimally by any of them] and therefore one
can generate squeezing in the atomic system, light system or both
independently (without performing measurements). On the other hand,
one can use $H_0$ to generate the swap operator, which (in the
Heisenberg picture) transforms
\begin{equation}\label{swap}
X_1 \leftrightarrow X_2, \quad P_1 \leftrightarrow P_2.
\end{equation}
This operation can be generated in a finite time. Thus, one can
use the interaction $H_0$ to realize a perfect interface between light
and atoms, which allows to use the atomic ensemble as a quantum memory
for light, as opposed to the case in Ref. \cite{ScJSP02} where this
result is obtained in the limit of very strong interaction.

\subsubsection{Optimal entanglement generation: infinitesimal case}

The problem that we consider now can be stated as follows. Let us
assume that we have some initial pure Gaussian state and we have
some interaction described by the general Hamiltonian
(\ref{generalHint}) at our disposal for a short time $\delta t$.
The initial state at time $t$ is described by some correlation
matrix of the form $\gamma(t)$ and possesses an entanglement
$E(t)$, where $E$ is some measure of entanglement. We would like
to increase the entanglement as much as possible.

Since for the case of two modes in a pure state there is a single
parameter that describes the entanglement [cf. \Eqref{Sstandard}], all
entanglement measures
are monotonically dependent on each other. One particular measure is
the parameter $r$ appearing in \Eqref{gamma2}, $\entm_0(\gamma)=r$. In
fact, $\entm_0$ is the log-negativity \cite{ViWe01} of the Gaussian
state. Thus, we have for every entanglement measure $\entm$:
$\entm(\gamma)=\entm(r)$. We use the obvious notation
$E(t)\equiv \entm[\gamma(t)]$ when considering the time-evolution of $\entm$.
Mathematically, our goal is to maximize the \emph{entanglement
rate} \cite{DuVCLP00}
\begin{equation}
\frac{d\entm}{dt}=\mbox{lim}_{\delta t \rightarrow 0} \frac{\entm(t+\delta
t)-\entm(t)}{\delta t}
\end{equation}
by using the fast local operations. We find the following result:
\begin{equation}
\left. \frac{d\entm}{dt}\right|_{\rm opt}= \left.\frac{d\entm}{dr}\right|_{r(t)}
\Gamma_E[\gamma,H].
\end{equation}
The function $\Gamma_E$, which genuinely contains the optimal
entanglement increase, is given by
\begin{equation}\label{optErate}
\Gamma_{E,\mathrm{opt}}[\gamma(t),H]=s_1 e^l-s_2 e^{-l},
\end{equation}
where $s_1,s_2$ characterize the given interaction Hamiltonian, while
$l$ is a parameter that only depends on the local squeezing
of our state and can be determined through the following relation
[using the notation of Eqs.~(\ref{gamma},\ref{gamma2})]
\begin{eqnarray}\label{locsqparam}
\cosh(2l)&=&\frac{\det(A)}{-2\det(C)} \tr(A^{-2} C
C^T)\\
&=& \frac{1}{2}\tr[(S_1^T S_1)^{-1} \sigma_z S_2^T S_2
\sigma_z].\nonumber
\end{eqnarray}
Note that there is no divergence as $\det C\to0$ as is seen by the
second expression in \Eqref{locsqparam} \citfn{fnLocSq}{%
Note that the $S_k$'s in \Eqref{gamma2} are uniquely defined only if
$\gamma$ is not a product state (i.e., iff $C\not=0$), cf.~[56]
Given a product state with CM $\tilde S_1\tilde
S_1^T\oplus \tilde S_2\tilde S_2^T$, the $S_k$ are defined only up to
local rotations $S_k = \tilde S_kO_k$. These $O_k$ can be chosen such
that $O_1(S_1^TS_1)O_1^T=\mathrm{diag}(\sigma_{1-},\sigma_{1+})$ and
$O_2(S_2^TS_2)O_2^T=\mathrm{diag}(\sigma_{2+},\sigma_{2-})$, where
$\sigma_{k+}=e^{r_k}\geq\sigma_{k-}=e^{-r_k}, r_k>0$ are the singular
values of $S_k^TS_k$. This local operation achieves the maximum
$\cosh(r_1+r_2)$ for the RHS in \Eqref{locsqparam} as given by von
Neumann's trace theorem \cite{HoJo94}.}.

Thus we see that the entanglement rate depends on the initial
local squeezing of the two modes as well as the angle between the
two locally squeezed quadratures, but it does not depend on the
entanglement of the state. Rewriting $\Gamma_{E,\mathrm{opt}}$ as
$(s_1-s_2)\cosh l+(s_1+s_2)\sinh l$ we see that some Hamiltonians
can produce entanglement even if there is no local squeezing
present in the state (which implies that $l=0$), while others
(notably the beam splitter with $s_1=s_2=1$) cannot.

Note that the rate goes to infinity as
local squeezing is increased, in contrast to the case of qubits.
Given a CM $\gamma$, there are typically local rotations that enhance
the entanglement rate.

From these results we conclude that if the goal is to create as much
entanglement as possible it is more efficient to squeeze the state
locally first (if possible) before using the interaction; in
particular, the use of squeezed light \cite{KuPo00} is advantageous
compared to coherent light \cite{DuSCZ00}.

\subsubsection{Optimal squeezing generation: infinitesimal case}

Now we consider the problem of optimal squeezing generation in the
same set-up as in the previous subsection. We take as a measure of
squeezing of a correlation matrix $\gamma$, $\cS=\cS(\sqm)$, any
monotonically increasing function of $\sqm$, where $\sqm$ is minus
the logarithm of the smallest eigenvalue of $\gamma$. We find
\begin{equation}\label{optSqrate}
\left. \frac{d\cS}{dt}\right|_{\rm opt}=
\left. \frac{dS}{d\sqm}\right|_{\sqm(t)}
g_S[\gamma(t)]C_S(H).
\end{equation}
$C_S(H)$ is the \emph{squeezing capability} of the Hamiltonian and
it is given by $s_1-s_2$, where the $s_i$'s are the restricted
singular values of $K$, given in (\ref{Kform}) and
\begin{equation}
g_S(\gamma)=2 \|\vec{x}_1\| \|\vec{x}_2\|\leq 1,
\end{equation}
quantifies how ``squeezable'' the state $\gamma$ is by interactions of
the type (\ref{generalHint}). Here $\hat{x}^T=(\vec{x}_1,\vec{x}_2)$,
with $\vec{x}_1, \vec{x}_2\in\RR^2$ is the normalized eigenvector
corresponding to the minimal eigenvalue of $\gamma(t)$.

\subsubsection{Optimal squeezing and entanglement: finite case}

Now we consider the situation in which we start with both modes in
the vacuum state and we have a Hamiltonian $H$ for a finite
time (as well as instantaneous local operations). We show that the
optimal way to create entanglement is to apply local instantaneous
operations flipping the $X$ and $P$ variables of both systems
periodically after small times $\delt$. After a finite time $t$ (and
for $\delt\to0$) this produces (up to local rotations) a two-mode
squeezed state, which is both optimally squeezed and entangled. In
particular, $\sqm(t) = (s_1-s_2)t$ and $\entm_0(t)=(s_1-s_2)t$.

We also show that it is not possible to increase the entanglement
using Gaussian measurements during the evolution. We consider a
system with CM $\gamma$ and ancilla systems in vacuum state. We
allow for linear passive interactions (described by a symplectic
and orthogonal matrix $O$) between one system and the ancillas and
show that a Gaussian measurement does neither increase the
squeezing nor the entanglement. This result implies that our
method is optimal even if we allow for feedback, something which
has been recently considered in the context of spin squeezing
generation \cite{ThMW02,BeSa02}.

For the case of atomic ensembles our result implies that there is a
method to improve the entanglement generation in present
experiments \cite{HaSSP99}.


\section{Simulation of interactions}

\label{SecSim}

In this section we characterize all the unitary evolutions which
we can generate within the given setup. That is we define the set
of unitary operators which can be written as (\ref{Simprot0}). The
first part of this section is devoted to the infinitesimal regime,
where we will in general derive the necessary and sufficient
conditions for Hamiltonian simulation. In the second part we are
concerned with the finite time regime. There we show that with
(almost) any Hamiltonian $H$ as in \Eqref{generalHint} and the local
operations corresponding to the Hamiltonians given in (\ref{Hloc}) it
is possible to generate any unitary gate.

\subsection{Method of Hamiltonian simulation}

A central result in the theory of Hamiltonian simulation
\cite{BeCLLLPV01} states that an alternating sequence of
manipulations and interactions as given in (\ref{Simprot}) is
equivalent to a fictitious free evolution due to a certain
effective Hamiltonian $H_{\mathrm{eff}}$, i.e. produces a unitary
transformation%
\[
\mathcal{U}=e^{-iH_{\mathrm{eff}}t^{\prime}}%
\]
and%
\begin{equation}
\kappa
H_{\mathrm{eff}}=\overset{n}{\underset{i=1}{\sum}}p_{k}\left(
\widetilde{V}_{i}^{\dagger}\otimes\widetilde{W}_{i}^{\dagger}\right)
H\left(
\widetilde{V}_{i}\otimes\widetilde{W}_{i}\right)  \label{Heff}%
\end{equation}
where $\kappa:=t^{\prime}/t$, $t:=\sum_{i=1}^{n}t_{i}$, the
$p_{k}:=t_{k}/t$ form a probability distribution and the
$\widetilde{V}_{i}\otimes\widetilde {W}_{i}$ follow uniquely from
the interspersed control operations $V_{j}\otimes W_{j}$ (and vice
versa). Obviously one can in this way \emph{simulate} an evolution
due to a Hamiltonian $H_{\mathrm{eff}}$ by means of a given
Hamiltonian $H$.

\Eqref{Heff} has a clear interpretation: A protocol proceeding in
infinitesimal time steps yields a mean Hamiltonian which is a
weighted sum of locally transformed variants of the original
Hamiltonian $H$. The so-called \emph{simulation factor} $\kappa$
is the ratio of simulated time $t^{\prime}$ and time of simulation
$t$ and, therefore, is a measure for the efficiency of the
simulation. The case $\kappa\geq1$ corresponds to the \emph{efficient
simulation}.

\subsection{Necessary and sufficient condition}

We associate to the general non--local interaction Hamiltonian
(\ref{generalHint}) the real $2\times2$ matrix $K$ as in
(\ref{Kform}). The action of a local rotation
$V(\varphi)=\exp[-i(X^{2}+P^{2})\varphi/2]$ on
the canonical operators $X$ and $P$ can be expressed by%
\begin{align}
V\left(
\begin{array}
[c]{c}%
X\\
P
\end{array}
\right)  V^{\dagger}  & =R\left(
\begin{array}
[c]{c}%
X\\
P
\end{array}
\right)  \quad\text{where}\nonumber\\
R=R(\varphi)  & =\left(
\begin{array}
[c]{cc}%
\cos\varphi & -\sin\varphi\\
\sin\varphi & \cos\varphi
\end{array}
\right)  \in SO(2,\mathbb{R}). \label{Rot}%
\end{align}
Thus we can associate to all local rotations $V_i,W_i$
(\ref{Hloc}) real orthogonal
$2\times2$ matrices $R,S\ldots$ with determinant $+1$. Consequently we have%
\begin{equation}
\left(  V\otimes W\right)  H\left(  V^{\dagger}\otimes
W^{\dagger}\right) =(X_{1},P_{1})R^{T}KS\left(
\begin{array}
[c]{c}%
X_{2}\\
P_{2}%
\end{array}
\right)  . \label{Transham}%
\end{equation}

Furthermore we use that for any matrix $K$ as given in
(\ref{Kform}) there exists a singular value decomposition
$K=OD\tilde{O}$ where $O,\tilde{O}\in O(2,\mathbb{R})$,
$D=\mathrm{diag}(\sigma_{1},\sigma_{2})$ and the singular values
$\sigma_{1}\geq\sigma_{2}\geq0$ of $K$ are unique. If we restrict
ourselves on \emph{special} orthogonal matrices we can still find
matrices
$R,S\in SO(2,\mathbb{R})$ such that%
\begin{equation}\label{Ksvd}
K=R\left(\begin{array}[c]{cc}
s_{1} & 0\\ 0 & s_{2}
\end{array}\right)  S
\end{equation}
and $s_{1}=\sigma_{1}$,
$s_{2}=\mathrm{sign}[\mathrm{\det}(K)]\sigma_{2}$ \cfn{notesign}.
Without loss of generality we may always assume that%
\begin{equation}\label{Resvals}
s_{1}\geq\left|  s_{2}\right|.
\end{equation}
Then these two values are uniquely defined and shall be called
\emph{restricted singular values} of $K$.

Assume now we want to simulate, in the above sense, some
Hamiltonian $H^{\prime}$ by means of some other Hamiltonian $H$,
both of the form (\ref{Kform}). Let $s_{1},s_{2}$ and
$s_{1}^{\prime},s_{2}^{\prime}$ denote their respective restricted
singular values. Then we have the following result:

$H$ can efficiently simulate $H^{\prime}$ iff%
\begin{equation}\label{Cond}
\begin{array}[c]{rcl}%
s_{1}+s_{2} & \geq & s_{1}^{\prime}+s_{2}^{\prime}\\
s_{1}-s_{2} & \geq & s_{1}^{\prime}-s_{2}^{\prime}.
\end{array}
\end{equation}
The proof is elementary but requires some effort in notation such
that we postpone it to \Aref{AppSim}.

\subsection{Discussion}

Since the number of relevant parameters characterizing an
interaction Hamiltonian is two, one can nicely illustrate the above
result:
\begin{figure}[ht]
\includegraphics[
trim=0.000000in 0.010360in 0.000000in -0.010360in, height=1.407in,
width=1.5489in
]%
{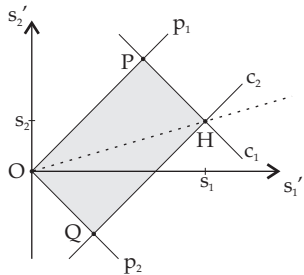}%
\caption{\label{FSim} Illustration of the accessible region in the
$(s_{1}^{\prime},s_{2}^{\prime})$-plane for the case
$s_{2}>0$. Coordinates of relevant points:
$H=(s_{1},s_{2}),$ $P=\frac{s_{1}+s_{2}}{2}(1,1),$
$Q=\frac{s_{1}-s_{2}}{2}(1,-1)$. See text for explanation.}
\end{figure}
The \Fref{FSim} illustrates the following geometrical relations:
Point $H=(s_{1},s_{2})$ denotes the original general Hamiltonian.
Lines $p_{1}$ and $p_{2}$ indicate the boundaries where
$s_{1}^{\prime}=\pm s_{2}^{\prime}$ respectively and are due to
premise $s_{1}^{\prime}\geq\left|  s_{2}^{\prime }\right|  $.
Lines $c_{1}$ and $c_{2}$ stem respectively from the first and
second inequality constituting the necessary and sufficient
condition. The region of accessible Hamiltonians, i.e. points
$H^{\prime}=(s_{1}^{\prime },s_{2}^{\prime})$ is thus contained in
the rectangle $OPHQ$. One can even visualize how this set deepens
with increasing time of simulation by parameterizing
$H(t)=(s_{1}t,s_{2}t)$. Thus, $H$ moves outward on the dashed line
while $P$ and $Q$ move on $p_{1}$ and $p_{2}$ respectively. It is
therefore just a matter of time to reach any point in the quadrant
enclosed by $p_{1}$ and $p_{2}$.

It is also quite instructive to consider certain special cases:
(i) For $s_{2}=s_{1}$ ($s_{2}=-s_{1}$) the dashed line coincides
with $p_{1}$ ($p_{2}$), respectively. This is a trivial case where
we are confined to simulate locally equivalent variants of the
original Hamiltonian (see \Aref{AppSim}). Therefore, Hamiltonians
whose restricted singular values are of equal modulus are nearly
useless for the purpose of Hamiltonian simulation. (ii) For
$s_{2}=0$ or, equivalently, $\det(K)=0$ the picture gets symmetric
with respect to the $s_{1}^{\prime}$-axis. This symmetrization can
be interpreted in terms of time efficiencies, as we shall explain
in the following.

Based on the criterion above one can ask for time efficiencies and
especially for \emph{time optimal protocols}. Time optimal
simulation is achieved if the simulation factor
$\kappa=t^{\prime}/t$ [see \Eqref{Heff}] gets maximal.
Without loss of generality we set $t^{\prime}=1$ such that
$\kappa=1/t$. Given now $H$ and $H^{\prime}$ with restricted
singular values $s_{1},s_{2}$ and $s_{1}^{\prime},s_{2}^{\prime}$
we can determine the minimal time of simulation as
$t_{\min}:=\underset{t}{\min}\{t:(s_{1}+s_{2})t\geq
(s_{1}^{\prime}+s_{2}^{\prime}),(s_{1}-s_{2})t\geq(s_{1}^{\prime}%
-s_{2}^{\prime})\}$. We find%
\begin{equation}\label{sim-tmin}
t_{\min}=\left\{
\begin{array}
[c]{c}%
\frac{s_{1}^{\prime}+s_{2}^{\prime}}{s_{1}+s_{2}}\\
\frac{s_{1}^{\prime}-s_{2}^{\prime}}{s_{1}-s_{2}}%
\end{array}
\right.  \quad\text{if\quad}%
\begin{array}
[c]{c}%
\frac{s_{2}^{\prime}}{s_{1}^{\prime}}\geq\frac{s_{2}}{s_{1}}\\
\frac{s_{2}^{\prime}}{s_{1}^{\prime}}<\frac{s_{2}}{s_{1}}%
\end{array}.
\end{equation}
Thus the efficiency of simulation depends strongly on whether
$\mathrm{sign}%
(s_{2}^{\prime})=\mathrm{sign}(s_{2})$ or not, the last case being
more time consuming. Only when $s_{2}=0$ [case (ii) above] it is
equally expensive (in terms of costs of interaction time) to
simulate either kind of Hamiltonians $H^{\prime}$
$[\mathrm{sign}(s_{2}^{\prime})\lessgtr0]$, a fact which is
reflected in the above mentioned symmetrization. Correspondingly,
the optimal time of simulation or, so to say, the \emph{minimal}
\emph{interaction costs}
\cite{ViHC01} are in this case uniquely determined by%
\begin{equation}
t_{\min}=(s_{1}^{\prime}+\left|  s_{2}^{\prime}\right|  )/s_{1}.
\label{Tmin}%
\end{equation}

\subsection{Application to $X_{1}X_{2}$-interaction}

Let us outline some conclusions out of this result for the
interaction $H=X_{1}X_{2}$. The restricted singular values of $H$
are obviously $s_{1}=1$ and $s_{2}=0$. Therefore we can
efficiently ($\kappa=1$, i.e. $t^{\prime}=t$) implement all
Hamiltonians $H^{\prime}$ whose restricted singular values
fulfill%
\begin{equation}
s_{1}^{\prime}+\left|  s_{2}^{\prime}\right|  \leq1.\label{Simcond}%
\end{equation}
Hence we can choose freely any interaction described by a
Hamiltonian $H^{\prime}$ satisfying (\ref{Simcond}) for the
purpose of creating entanglement or squeezing as we will do in the
following sections.

As an example as well as to give a basis for further results we
shall consider here two kinds of well known unitary
transformations: the \emph{beam-splitter}
operator%
\begin{equation}\label{Bs}
U_{\mathrm{bs}}(t):=e^{-iH_{\mathrm{bs}}t}\quad\text{where}\quad
H_{\mathrm{bs}}=X_{1}P_{2}-P_{1}X_{2}%
\end{equation}
and the \emph{two-mode squeezer}%
\begin{equation}\label{Tms}
U_{\mathrm{tms}}(t):=e^{-iH_{\mathrm{tms}}t}\quad\text{where}\quad
H_{\mathrm{tms}}=X_{1}X_{2}-P_{1}P_{2}.
\end{equation}

As mentioned already, the action of $\Ubs(\pi/2)$ corresponds to
swapping the states of the first and the second mode, i.e. it
transforms $X_{1,}P_{1}\rightarrow -X_{2},-P_{2}$ and
$X_{2,}P_{2}\rightarrow X_{1},P_{1}$. Note that the global phase
thereby acquired by subsystem 1 can be corrected locally.

Application of $U_{\mathrm{tms}}(t)$ squeezes the EPR modes
$(X_{1}+X_{2})$ and $(P_{1}-P_{2})$ by a factor $e^{-2t}$ and
therefore also entangles the two systems, as we shall see.

In order to perform these operations by means of the
$X_{1}X_{2}$-interaction we have to determine the restricted
singular values of $H_{\mathrm{bs}}$ and $H_{\mathrm{tms}}$. One
finds for $H_{\mathrm{bs}}$ $s_{1}=1,s_{2}=1$ and for
$H_{\mathrm{tms}}$ $s_{1}=1,s_{2}=-1$. Since in both cases
condition (\ref{Simcond}) is not met we cannot \emph{efficiently}
simulate these Hamiltonians. But nevertheless we can determine
strategies for infinitesimal simulations being time optimal. The
minimal time of simulation can be calculated using (\ref{Tmin})
and yields a maximal simulation factor $\kappa=1/t_{\min}=1/2$ for
both, the beam-splitter and the squeezer. Thus, in order to
implement $\Ubs(t^\prime)$ we need at least a time $t=2t^\prime$
and to create squeezing by a factor $e^{-2t^\prime}$ it will take
a time $2t^\prime$, i.e. to implement $U_\mathrm{tms}(t^\prime)$ we need
a time $t=2t^\prime$. Explicit simulation protocols can be
constructed following \Aref{AppSim}.

\subsection{Simulation of unitary operators and state engineering}

Until now we have focused on the regime of infinitesimal times in
order to clarify which \emph{unitary evolutions} we can simulate
by means of the given interaction. We found that we can do so --
more or less efficiently -- for all evolutions governed by
Hamiltonians of the form (\ref{Kform}), but no more. This leaves
open the question which \emph{unitary operations} can in general,
i.e., for finite times, be realized with a given interaction and
local rotations.

As we show in the following, any interaction described by some
Hamiltonian $H$ where $s_{1}\neq \left| s_{2}\right| $ together
with local rotations is sufficient to realize \emph{any unitary
operation} of the form $\exp (iG)$, where $G$ is a quadratic
expression in the operators $X_{k},P_{k}$. That is, any Gaussian
unitary transformation of the two modes can be obtained. This
implies, that any desired pure Gaussian state can be
``engineered'' starting from any given (pure Gaussian) input
state.

As we show in \Aref{AppGSim}, any $U=\exp (-iG)$ can be decomposed as
\begin{equation}\label{Decomp}
\parbox{\columnwidth}{\begin{eqnarray*}
U &=&\left( V_{5}\otimes
W_{5}\right)U_{\mathrm{bs}}(t_{5})\left(V_{4}\otimes W_{4}\right)
\times \nonumber\\
&&\times U_{\mathrm{tms}}(t_{4})\left(
V_{3}\otimes W_{3}\right) U_{\mathrm{bs}}(t_{3})\left( V_{2}\otimes
W_{2}\right)U_{\mathrm{tms}}(t_{2})\times\\
&&\times\left( V_{1}\otimes W_{1}\right) U_{\mathrm{bs}}(t_{1})\left(
V_{0}\otimes W_{0}\right),\nonumber
\end{eqnarray*}}
\end{equation}
where all $\left( V_{i}\otimes W_{i}\right) $ are local rotations,
$U_{\mathrm{bs}}(t_{i})$ is a beam-splitter and
$U_{\mathrm{tms}}(t_{i})$ a two-mode squeezing operation as
defined in Eqs. (\ref{Bs}) and (\ref{Tms}). Since all Hamiltonians
with $s_{1}\not= \left| s_{2}\right| $ can be used to simulate
beam-splitters and two-mode squeezers one can reach any desired
unitary $U$ and therefore also any desired Gaussian state.


\section{Entanglement and Squeezing}\label{SecEnt}

In the previous section we characterized the time-evolutions on
the joint system which can be realized using a given interaction
Hamiltonian of the form (\ref{generalHint}) and the control
operations provided by \Eqref{Hloc}. In this section we determine
the optimal way to use these tools for the generation of
entanglement and squeezing between the two subsystems in both, the
infinitesimal and the finite regime.

Our derivations make extensive use of the formalism of Gaussian
states and operations. The necessary concepts and notation are
introduced in section \ref{notation} and then put to work in the cases
of infinitesimal (\ref{SecRate}) and finite (\ref{SecFinTim})
times.

\subsection{State Transformations and Measures of Entanglement and
Squeezing}\label{notation}

We show here how Gaussian states evolve under a general quadratic
Hamiltonian and then introduce some entanglement and squeezing
measures for Gaussian states.

\subsubsection{State Transformation}
A quadratic interaction Hamiltonian (\ref{generalHint})
characterized by a matrix $K$ as in \Eqref{Kform}
generates a linear time-evolution of the $X$ and $P$
operators. Solving the Heisenberg equations for $\vec R =
(X_1,P_1,X_2,P_2)^T$ we find
\begin{equation}\label{integrate}
\vec{R}(t)=e^{Mt}\vec{R}(0)=S(t) \vec{R}(0),
\end{equation}
where
\begin{equation}\label{defM}
M=\left(\begin{array}{cc} 0&L\\ \tilde{L}&0 \end{array}\right),
\end{equation}
with
\begin{equation}\label{defL}
L= \left(\begin{array}{cc} c&b\\ -a&-d \end{array}\right)=J^TK,
\mbox{ and } \tilde{L}= -JL^TJ^T=J^TK^T,
\end{equation}
where
\begin{equation}\label{J}
J=\left( \begin{array}{cc} 0&-1\\1&0 \end{array} \right).
\end{equation}
Note that for $0\not=-\det(L)=: \alpha$ we have $\tilde{L}=\alpha L^{-1}$.
Using the fact that $M^2=\alpha \one$ we can easily re-express
\Eqref{integrate} and find
\begin{equation}\label{time-ev}
S(t)=\cosh(\sqrt{\alpha}t) \one +\sinh(\sqrt{\alpha}t) / \sqrt{\alpha}M.
\end{equation}

Thus, every evolution generated by a Hamiltonian (\ref{generalHint})
is uniquely characterized by a symplectic transformation $S(t)$ of the
form (\ref{time-ev}). Note that any such transformation can be
written in its \emph{standard form}
\begin{equation}\label{Sstandard}
S(t)=\cosh(\sqrt{\alpha} t)(O_1\oplus O_2) \left(\begin{array}{cccc} 1&0&h_1&0\\
0&1&0&-h_2\\
h_2&0&1&0\\
0&-h_1&0&1 \end{array}\right)(O_1\oplus O_2)^T,
\end{equation}
where $O_1,O_2 \in SO(2,\RR)$ perform the restricted singular
value decomposition of $L$, and $h_k=\tanh(\sqrt{\alpha
t})/\sqrt{\alpha}s_k$, where $s_k$ are the restricted singular values
of $L$, which clearly coincide with those of $K$. In particular the
Hamiltonian $H_0=X_1X_2$ of \Eqref{Hint} generates an time-evolution
described by the symplectic matrix
\begin{equation}\label{St}
S_0(t) = \left( \begin{array}{cccc}
1&0&0&0\\
0&1&-t&0\\
0&0&1&0\\
-t&0&0&1
\end{array} \right),
\end{equation}
i.e. $\alpha=0, (s_1,s_2)=(1,0)$, and $O_1=J$ [see (\ref{J})] and
$O_2=-\id$.

In the Schr{\"o}dinger picture a linear time-evolution as in
(\ref{integrate}) transforms the CM $\gamma$ as
\begin{equation}\label{time-evCM}
\gamma(t) = S(t)\gamma S(t)^T.
\end{equation}
In the next subsection we address the case of very short interaction
time, i.e., we consider $S(\delta t)$ for an infinitesimally short
time step $\delta t$. In this case we obtain
\begin{equation}\label{Sdt}
S(\delta t)=\one+\delta t M,
\end{equation}
and the correlation matrix $\gamma(t)$ transforms to first order as
\begin{equation}\label{gammadt}
\gamma(t+\delta t)=\gamma(t)+\delta t[M \gamma(t)+\gamma(t) M^T].
\end{equation}

Let us in the following write the $4\times4$ CM of the two-mode
Gaussian state as a block matrix as in \Eqref{gamma}
with $2\times2$ matrices $A,B,C$. Then $A$ refers to the first
system and is the CM belonging to the reduced density operators of
the system $1$. Note that for all CMs $\det(\gamma)\geq1$, and
equality holds if and only if (iff) the state is pure. Since our
initial state is pure and we consider unitary transformations (and,
later, complete Gaussian measurements) this implies that we are only
concerned with pure states at all times.

\subsubsection{Entanglement and Squeezing of Gaussian
States}

As one can see in equation (\ref{gamma2}), the single parameter
which characterizes the non-local properties of a pure state is
the two-mode squeezing parameter $r$. This automatically implies that any
monotonic function of this parameter can be used to quantify the
entanglement of pure Gaussian two-mode states and we are free to
choose \citfn{noteErate}{The canonical measure of entanglement for
pure states is the entropy of entanglement $E$, i.e., the von
Neumann entropy of the reduced state. For pure Gaussian states it
is $E(\ket{\psi})=\cosh(r)^2 \log[\cosh(r)^2]-\sinh(r)^2
\log[\sinh(r)^2]$, where $r=[\acosh(\sqrt{\det A})]/2$, with $A$ the
CM of the reduced state \cite{vEnk99}. Consider now any function
$f(r)$ such that $E(f)$ is a monotonic function of $f$. The
maximization of the rate of $E$ with respect to the evolution is
then equivalent to the maximization of the rate of $f$. The reason
for this is that $\mbox{max}_H
\left.\left(\frac{dE}{dt}\right)\right|_{t_0}=\mbox{max}_H
[\left.\left(\frac{dE}{df}\right)\right|_{t_0}
\left.\left(\frac{df}{dt}\right)\right|_{t_0}]=
\left.\left(\frac{dE}{df}\right)\right|_{t_0} \mbox{max}_H
\left.\left(\frac{df}{dt}\right)\right|_{t_0}$. Since $E$ is a monotonic
function we have that $\left.\left(\frac{dE}{df}\right)\right|_{t_0} >0$,
which implies that maximizing $\left.\left(\frac{dE}{dt}\right)\right|_{t_0}$
with respect to the evolution is equivalent to maximize
$\left.\left(\frac{df}{dt}\right)\right|_{t_0}$ respect to the evolution.} the most
convenient measure.

One such quantity is $E_p(\gamma)= \det A =\cosh(r)^2$, the
determinant of the CM corresponding to the
reduced density.  It is related to the \emph{purity} of the
reduced density matrix \citfn{notepurity}{In
general the purity is not a measure of entanglement, but for pure
states, $\ket{\psi}$ the purity $\tr_2(\rho_{red}^2)$, where
$\rho_{red}=\tr_1(\proj{\psi})$, decreases the more entangled
$\ket{\psi}$ is. Therefore we may use, e.g., the inverse square of
purity, i.e., $\cP(\ket{\psi}) = [\tr(\rho_\mathrm{red}^2)]^{-2}$
to quantify how entangled a given pure state is.\newline
For a general two--mode Gaussian
state with CM $\gamma$ as in \Eqref{gamma} tracing over the second
system yields a reduced density matrix which is Gaussian with CM
$\gamma_\mathrm{red} = A$.  The purity of the reduced state is
therefore given by $\det(A)$ as \cite{Scut98}
$\cP(\gamma)=\{\tr[\rho_{\mathrm{red}}(\gamma)^2]\}^{-2}=\det A$.}.
As mentioned before, the determinant of a CM is one, iff the state is
pure, which implies that $E_p(\gamma)=1$ iff the state is not
entangled, i.e., iff $r=0$.

For the last part of this section another measure of entanglement,
namely the \emph{negativity} $\cN$ introduced in Ref.\
\cite{ViWe01} is most convenient to use. For a $1\times1$ Gaussian
state with CM $\gamma$ the negativity is given by the inverse of
the smallest symplectic eigenvalue of the partially transposed CM
$\tilde\gamma=\Lambda\gamma\Lambda$, which can easily be
calculated \cite{ViWe01} as
\begin{equation}\label{neg}
\cN(\gamma) =
\left[\mbox{min}\{\mbox{sing.val.}\left(J_2^T\gamt J_2\gamt
\right)\}\right]^{-1/2}.
\end{equation}
Here $\Lambda$ is the $4\times4$ diagonal matrix $\diag(1,1,1,-1)$
(which implements partial transposition, see \cite{Simo99}) and $J_2=J\oplus
J$ is the symplectic matrix for two modes.

The other interesting quantity that characterizes Gaussian states besides the
entanglement is the \emph{squeezing} inherent in the state, i.e., by
how much the variance of some (passive-linearly transformed)
quadrature is reduced below the standard quantum limit. The reduced
variance is given by the smallest eigenvalue $\sev(\gamma)$ of
$\gamma$ and we define the squeezing of a state with CM $\gamma$ as
the inverse of $\sev(\gamma)$
\begin{equation}\label{Squeez}
\cS(\gamma) =
\mbox{min}\{\mbox{eig}(\gamma)\}^{-1}=[\sev(\gamma)]^{-1}.
\end{equation}
In a situation like the one we consider here where only orthogonal
operations are freely available, the squeezing of a state
represents a valuable resource which can be used, e.g., for the
creation of entanglement \cite{WoEP02} and which should be created
as efficiently as possible.

\subsection{Optimal Entanglement/Squeezing Rates}\label{SecRate}

The goal of this section is to determine the optimal strategy for
the generation of entanglement [squeezing] in an (infinitesimally)
small time step $\delta t$. That is, given a pure Gaussian state
$\rho$ with CM $\gamma$ and an interaction Hamiltonian $H$ as in
\Eqref{generalHint} we look for the best choice of the local
rotations $V\otimes W$ such that $e^{-iH\delta t}(V\otimes
W)\rho(V\otimes W)^\dagger e^{iH\delta t}$ is as entangled
[squeezed] as possible. Stating this problem mathematically: We
maximize the \emph{entanglement [squeezing] rate}, that is the
time-derivative of the chosen entanglement [squeezing] measures
$E$ [$\cS$] under the time-evolutions obtainable in the given
setting.

\subsubsection{Maximizing the Entanglement Rate}
As measure of entanglement we use $\entm_0$, where $\entm_0(\gamma)$ is the
two-mode squeezing parameter $r$ \footnotemark[\value{noteErate}]
defined in (\ref{gamma2}). The entanglement rate is then simply given by
\begin{equation}\label{Gamma_E}
\Gamma_E=\left.\frac{d\entm_0}{dt}\right|_{t=0} =\lim_{\delta
t\rightarrow 0} \frac{r(\delta t)-r}{\delta t},
\end{equation}
where $r\equiv r(0)$ is the entanglement of the initial CM $\gamma$.

In order to determine $\Gamma_E$ we use, following
\Eqref{Gamma_E}, the formula $\Gamma_{E_p}=\sinh(2r)
\Gamma_E=2\sqrt{-\det(A)\det(C)}\Gamma_E$, where $\Gamma_{E_p}$
denotes the entanglement rate corresponding to the purity-related
measure $E_p$.

Let $H$ as in \Eqref{Kform} be the given Hamiltonian. It generates
an evolution given by the symplectic transformation
$\bar{S}(\delta t)$, which we write in its standard form
(\ref{Sstandard}) as $\bar{S}(\delta t) := (\bar{O}_1\oplus
\bar{O}_2) S(\delta t) (\bar{O}_1\oplus \bar{O}_2)^T$. Since local
operations cannot increase the entanglement the only way in which
the local control operations may help is to rotate the state by
$\tilde O_1\oplus \tilde O_2$ before applying $H$. Thus the best
strategy yields a $\gamma(\delta t)$ that can be written as
\begin{equation}\label{evolutiongamma}
\gamma(\delta t) = S(\delta t)(O_1\oplus O_2)\gamma (O_1\oplus O_2)^T
S(\delta t)^T,
\end{equation}
where we defined $O_i:=\bar{O}_i^T\tilde O_i$ and omitted the
irrelevant final local rotations coming from $\bar{S}(\delta t)$.
Writing $\gamma(\delta t)$ in the form (\ref{gamma}) and using
\Eqref{gammadt} it is straight forward to determine the CM
corresponding to the reduced state,
\begin{equation}
A(\delta t)=O_1 A O_1^T + \delta t (L_0 O_2 C^T O_1^T+\, \mathrm{H.c.}),
\end{equation}
where $L_0=\diag(s_2,-s_1)$ is determined by the Hamiltonian $H$,
cf. \Eqref{Sstandard} and \Eqref{defM}. One quickly sees that
$\det[A(\delta t)]=\det(A)[1+2 \delta t\, \tr(L_0O_2 C^T A^{-1}
O_1^T)]$, where we used the simple  relation for $2\times 2$
matrices: $\det(X+\delta t Y)=\det(X)[1+\delta t\, \tr(X^{-1}
Y)]+o(\delta t^2)$ and the fact that $A$ is symmetric and
invertible.

For the entanglement rate corresponding to $E_p$ we obtain
$\Gamma_{E_p}=2 \det(A) \tr(L_0O_2 C^T
A^{-1}O_1^T)$. As mentioned before we can from this easily determine the
rate $\Gamma_E$ corresponding to the two-mode squeezing parameter
namely we have
\begin{equation}\label{ErateE}
\Gamma_{E}=\sqrt{\frac{\det(A)}{-\det(C)}} \tr(L_0O_2
C^T A^{-1} O_1^T) = \tr(L_0O_2YO_1^T),
\end{equation}
where we have defined $Y := \sqrt{\det(A)/[-\det(C)]} C^T A^{-1}$.

Our aim is to maximize this expression with respect to the special
orthogonal matrices $O_1$ and $O_2$. Note that $\det Y = -1$,
which can be easily verified using \Eqref{gamma2}. Therefore $Y$
has the restricted singular values $e^l,-e^{-l}, l\geq 0$. Using
that $L_0$ is diagonal it is straight forward to verify that the
maximum of \Eqref{ErateE} is achieved when choosing $O_1, O_2$
such that they diagonalize $Y$  such that $O_2Y
O_1^T=\diag(e^l,-e^{-l})$.  Then the optimal choice for
$\tilde{O}_i$ is
\begin{equation}
\tilde{O}_{i,\mathrm{opt}} = \bar{O}_i O_i,
\end{equation}
with $\bar{O}_i$ given by $\bar{S}(\delta t)$. The best state to
let $H$ act on is thus
$\gamma_\mathrm{opt}=(\tilde{O}_{1,\mathrm{opt}}\oplus
\tilde{O}_{2,\mathrm{opt}})\gamma(\tilde{O}_{1,\mathrm{opt}}\oplus
\tilde{O}_{2,\mathrm{opt}})^T$. Note that $l$ which determines the
singular values of $Y$ can be easily determined by
\Eqref{locsqparam} \cfn{fnLocSq}.

In summary, given an interaction Hamiltonian $H$ corresponding to a
matrix $K$ and an initial state with CM $\gamma$ the optimal state
preparation by local rotations (before letting $H$ act) can be
understood as a two-step procedure. First transform $\gamma$ locally
such that $C^TA^{-1}$ is diagonal [restricted singular value
decomposition, cf. \Eqref{Ksvd}]. If $K$ was already in its restricted
singular value decomposition, we are
done. Otherwise the second step of the state preparation can be
viewed (in the Heisenberg picture) as the restricted singular value
decomposition of $K$.
Then the optimal entanglement rate (entanglement is measured
by $\entm_0$) is given by \Eqref{optErate} in terms of the singular
values $s_k$ of the Hamiltonian matrix $K$ and the local squeezing
parameter $l$ of the given state $\gamma$.

\begin{figure}
\includegraphics[scale=0.9]{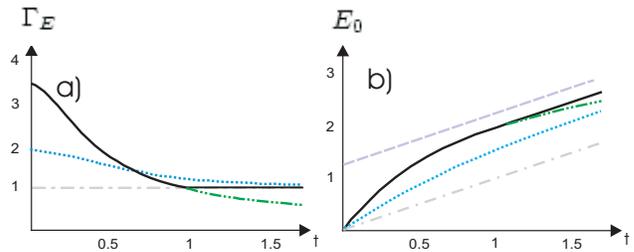}
\caption{\label{FRate} (a) The entanglement rate obtained for the
squeezed state $\gamma_\mathrm{in}$ (\ref{gamsqu}) as initial state and
various strategies. The solid line represents the optimal-rate
strategy derived in this section; the dotted line represents the rate
obtained by simulating the two-mode squeezing Hamiltonian
$H_\mathrm{tms}$; the ``dot--dot--dashed'' line 
represents the rate obtained for the natural Hamiltonian
$H_0=X_1X_2$. For the vacuum state as initial state we obtain the
constant rate $1$ (dashed line)  (b) The entanglement
created by the different strategies [same styles as in a) for the different
scenarios]. The dashed line represents the upper bound 
\Eqref{bound}.}
\end{figure}

In the \Fref{FRate} we compare the entanglement rates and the
entanglement obtained for different strategies using the ``natural
Hamiltonian'' $H_0$. As initial state we
consider the product of the vacuum state in the first system and
the squeezed vacuum in the second system, i.e.,
\begin{equation}\label{gamsqu}
\gamma_\mathrm{in}=\id_2\oplus
\left(\begin{array}{cc}
e^{-r}&0\\
0&e^r
\end{array}\right),
\end{equation}
with squeezing parameter $r=2.5$. We compare the strategy in which the
rate of entanglement creation is optimized at each time to two
simpler ones, namely to just apply the natural Hamiltonian $H_0$ 
or to simulate the two-mode squeezing Hamiltonian
$H_\mathrm{tms}=X_1X_2-P_1P_2$ using the optimal scheme of
\Sref{SecSim}. The rate-optimization 
strategy leads in fact to combination of the other two: 
one applies first the natural Hamiltonian for a finite time and then
(when the ``local squeezing'' $l$ has all been converted to two-mode
squeezing) one simulates $H_\mathrm{tms}$. Having initially 
local squeezing available clearly helps with entanglement generation:
for an initial unsqueezed state the optimal rate is constant
$\Gamma_E=1$ . 

Fig.~\ref{FRate}b shows that the optimization strategy can lead to
noticably more entanglement in the resulting state after finite time:
when the entanglement rate is optimized at each point, more
entanglement is produced than, e.g., with the interactions $H_0$ or 
$H_\mathrm{tms}$. However, optimizing the rate is in general not the best
strategy for the creation of entanglement, see \Fref{Fcounter}. 

\subsubsection{Maximizing the squeezing rate} \label{Subsecsqe}

As in the previous section we are given an interaction Hamiltonian
of the form (\ref{generalHint}), an initial Gaussian state with CM
$\gamma$, and we consider the case of infinitesimal interactions.
Our goal is here to determine for each $H$ and $\gamma$ the
strategy which maximizes the squeezing rate. We measure squeezing
by $\sqm(\gamma)=\log[\cS(\gamma)]$, where $\cS$ was defined in
\Eqref{Squeez} as the inverse of the smallest eigenvalue of
$\gamma$. The rate we are interested in is
\begin{eqnarray}\label{SqRate}
\Gamma_S & = &
\left.\frac{d}{dt} \log\cS[\gamma(t)]\right|_{t=0}\\
& =&
\frac{-1}{\sev(\gamma)}\lim_{\delta t\to0}\frac{\sev[\gamma(\delta
t)]-\sev(\gamma)}{\delta t}. \nonumber
\end{eqnarray}
Note that we use the logarithm of $\cS$ instead of $\cS$ for
convenience. It simplifies the formulas but since $\log$ is a
monotonic function maximizing the rate of $\log\cS$ implies a
maximal rate for $\cS$ as well \footnotemark[\value{noteErate}].

After applying the general strategy to the input
state with CM $\gamma$ we obtain $\gamma(\delta t)$ as in
\Eqref{gammadt}. Doing first order perturbation theory we
find that $\lambda_{\mathrm{min}}[\gamma(\delta
t)]=\lambda_{\mathrm{min}}(\gamma)+\delta t \hat x^T(M^T \gamma +
\gamma M)\hat x=\lambda_{\mathrm{min}}[1+\delta t \hat x^T(M^T+ M)\hat
x]$, where $\hat x$ is the normalized eigenvector corresponding to the
smallest eigenvalue $\lambda_\mathrm{min}(\gamma)$ of $\gamma$.  We
obtain for the squeezing rate:
\begin{equation}
\Gamma_S=\frac{-1}{\sev(\gamma)}[\hat x^T(M^T+ M)\hat x],
\end{equation}
which is maximized when $-\hat x^T(M^T+ M)\hat x$ is as large as
possible. Note that
\begin{equation}
M^T+M\equiv\left(\begin{array}{cc}
0&N\\ N^T&0
\end{array}\right),
\end{equation}
where $N=\tilde{L}+L^T=J^TK^T+K^{T}J$,
where $J$ is the $SO(2)$--matrix of \Eqref{J} and we have used the
definitions (\ref{defL}) and (\ref{Kform}). One quickly sees that
$N=N^T$. Writing $K$ in its restricted singular value decomposition
$K=SK_0R$, where $S,R\in SO(2,\RR)$ and
$K_0=\diag(s_1,s_2)$ as in \Eqref{Ksvd}, and using that $R,S$ commute
with $J$ we see that $N= R^T(J^TK_0+K_0J)S^T=C_S(H) R^T J^T\sigma_z
S^T$, where
\begin{equation}
C_S(H)=s_1-s_2
\end{equation}
is the \emph{squeezing capability} of the Hamiltonian $H$. Note that the
matrix $\tilde{O}:= R^T J^T\sigma_z S^T$ is orthogonal with
$\det(\tilde{O})=-1$ and that we can obtain any such $\tilde{O}$
choosing $R, S\in SO(2,\RR)$, i.e., by the local operations
applied to the initial state. Using the notation
$\hat x^T=(\vec x_1^T,\vec x_2^T)$, where $\vec x_1, \vec
x_2\in\RR^2$, we find $\Gamma_S=2 C_S(H)\vec x_1^T \tilde{O} \vec x_2 \leq 2
C_S(H) \mathrm{max}_{\tilde{O}} |\vec x_1^T \tilde{O} \vec x_2| = 2 C_S(H)
\|\vec x_1\| \|\vec x_2\|$, which gives an upper bound
\[\Gamma_S\leq 2 C_S(H) \|\vec x_1\| \|\vec x_2\|
\]
for $\Gamma_S$. This maximum can be reached for
$\tilde{O}_\mathrm{opt}$ such that $(-\tilde{O}_\mathrm{opt}\vec
x_2) || \vec x_1$. Given $\gamma$ (i.e., $\vec x_1, \vec x_2$) we
can calculate $\tilde{O}_\mathrm{opt}$ with
$\det\tilde{O}_\mathrm{opt}=-1$ which satisfies this condition.
This then determines the optimal choice of $R, S\in SO(2,\RR)$,
i.e. how to transform the initial state with CM $\gamma$ before
letting $H$ act in order to maximize the squeezing rate. One
simple choice yielding $\tilde{O}=\tilde{O}_\mathrm{opt}$ is
$S=\one$, i.e. nothing has to be done on the second and
$R_\mathrm{opt} = J\sigma_z \tilde{O}_\mathrm{opt}\in SO(\RR,2)$.
Thus, the optimal input state is given by $\gamma_\mathrm{opt}=(
R_\mathrm{opt}^T \oplus \one)\gamma ( R_\mathrm{opt} \oplus
\one)$.

In summary, we have shown that the maximal squeezing rate is
given by \Eqref{optSqrate} as a product of the squeezing capability
$C_S(H)$ of the given Hamiltonian and the squeezability $g_S(\gamma)$
of the given state. The optimal CM to let $H$ act on is $\gamma_\mathrm{opt}
=(R_\mathrm{opt}^T \oplus \one)\gamma (R_\mathrm{opt} \oplus \one)$, where
\begin{equation}
R_\mathrm{opt} = J^T\sigma_z\tilde{O}^T
\end{equation}
and $-\tilde{O}_\mathrm{opt}$ parallelizes $\vec x_1$ and $\vec x_2$.
Note that the fact that $\hat{x}$ is normalized implies that
$\Gamma_S \leq C_S(H)$ for any input state. Since we look at the
logarithm of the squeezing this implies that $\frac{d \cS(\gamma)}{dt}
\leq \cS(\gamma) C_S(H)$.


\subsection{Optimal entanglement generation from the vacuum
state}\label{SecFinTim} In practice, we are interested in creating
the largest amount of entanglement when $H$ acts for a
\emph{finite} total time $t$. Optimizing the rate of entanglement
creation at each time does lead to a local but not necessarily, as
we saw, the global maximum of the entanglement at time $t$
\cite{KrCi00}.

We now show how to employ the interaction $H$ to create the most
entanglement in a given time $t$. To this end, we make use of the
\emph{squeezing} of $\gamma$ which was introduced in
\Eqref{Squeez} as the smallest eigenvalue of $\gamma$. The
squeezing of $\gamma$ is known \cite{WoEP02} to give an upper
bound for the amount of entanglement of $\gamma$, with
$\cN(\gamma)\leq \cS(\gamma)$. We proceed as follows: First we
calculate the strongest squeezing that can be achieved after time
$t$. This also gives an upper bound for the entanglement that can
be obtained during this time.  Then we point out a strategy that
achieves the optimal squeezing and at the same time the strongest
entanglement compatible with the given squeezing, thus being
optimal on both counts.

The squeezing capability of a symplectic map $S$, i.e., the
factor by which the squeezing in a CM can be increased by the
application of $S$, is given by the inverse square of the smallest
singular value of $S$, since $\cS(S\gamma
S^T)\leq[\ssv(S)]^{-2}\cS(\gamma)$. Here and in the following we
use that for the smallest singular value of a product $AB$ we have
$\ssv(AB)\geq\ssv(A)\ssv(B)$. 
Now consider the symplectic map $S(t)$ corresponding to the unitary
evolution generated by an interaction Hamiltonian $H$ after time $t$, cf.
\Eqref{time-ev}. The singular values of $S(t)$ can easily be
calculated analytically. We need them only for small times to
first order in $t$, in which case we find:
\begin{equation}\label{singvalSt}
\sigma_\pm[S(t)] = \sqrt{1\pm \frac{1}{2}(s_1-s_2)t}+o(t)^2,
\end{equation}
where $s_1,s_2$ are the restricted singular values of the matrix $K$
[cf. \Eqref{Kform}] corresponding to $H$.

Since $S(t)=S(t/2)S(t/2)=\Pi_{k=1}^NS(t/N)$ we see immediately that 
$(\ssv[S(t)])^2\geq e^{-(s_1-s_2)t}$, which implies that the squeezing
capability of $S(t)$ is bounded by $e^{(s_1-s_2)t}$. Now consider a
strategy as in \Eqref{Simprot0}, alternating the use of $H$ for time
$t_k$ with local rotations $V_k\otimes W_k$. 
Note that the $t_k, k=1,\dots,N$, which sum to $t$, are not assumed to
be infinitesimal. The time-evolution effected by this strategy is
described by a symplectic map
\begin{equation}\label{strat}
S(t)=\Pi_k\tilde S_k,
\end{equation}
where $\tilde S_k=O_kS(t_k)O'_k$ and $O_k,O'_k$ are the local
rotations corresponding to $V_k\otimes W_k$. Clearly,
$\ssv[S(t)]\geq\Pi_ke^{-(s_1-s_2)t_k/2}=e^{-(s_1-s_2)t/2}$. Hence
$\cS[S(t)S(t)^T]\leq e^{(s_1-s_2)t}$, i.e. we have an
upper bound to the amount of squeezing that can be produced from an
initially unsqueezed pure state by applying $H$ for a total time $t$.

A strategy to achieve this optimum is the following: we choose the
local rotations $V_k,W_k$ as $\pi/2$-rotation in system 1 and $3\pi/2$
in system 2, the times $t_k$ all equal, and consider the limit $t_k\to0$. 
This corresponds to the situation considered in \Sref{SecSim} and
simulates the Hamiltonian related to $K' = (K+JKJ)/2$. 
Let $K=O_1\,\mathrm{diag}(s_1,s_2)O_2$, then we have that
$K'=1/2\,O_1[\mathrm{diag}(s_1,s_2)+\mathrm{diag}(-s_2,-s_1)]O_2$,
since rotations commute with $J$. That is, apart from local rotations
the strategy, which simulates the two-mode squeezing Hamiltonian with an
efficiency $(s_1-s_2)/2$, which is the optimal factor according to
\Eqref{sim-tmin}. Letting $H_\mathrm{tms}$ act for a time
$t'=t(s_1-s_2)/2$ (using up an interaction time $t$) transforms the
vacuum state into the two-mode squeezed state with CM
\begin{equation}\label{tmsCM}
\gamma_\mathrm{tms}(t') =
\left( \begin{array}{cc} \cosh 2t' \id&\sinh 2t'\sigma_z\\
\sinh 2t'\sigma_z&\cosh 2t'\id
\end{array} \right).
\end{equation}
which saturates the bounds derived above, since
$\cS[\gamma_\mathrm{tms}(t')]=e^{(s_1-s_2)t}$. 

Now we show that $\gamma_\mathrm{tms}$ in \Eqref{tmsCM} is also the
most entangled state that can be obtained after letting $H$ act for a
total time $t$. Using \Eqref{neg} for the negativity of a Gaussian
state with CM $\gamma=S(t)S(t)^T$ (i.e. an arbitrary strategy applied
to the vacuum state) we get
\[
\cN(\gamma) = [\cS(J^T\gamt
J\gamt)]^{-1/2}\leq\cS(\gamt)=\cS({\gamma})=e^{(s_1-s_2)t}.
\]
Since $\cN[\gamma_\mathrm{tms}(t')]=e^{(s_1-s_2)t}$ the simulation of
two-mode squeezing is the optimal strategy for both squeezing and
entanglement generation. Note that even a rough approximation of the
optimal strategy, i.e., a strategy consisting of just two or three
steps already yields a marked improvement in generated squeezing and
entanglement.

Up till now we have only considered the unitary evolution of the
initial state. There are, however, further tools available in
current experiments. There might be additional light modes
(ancillas) in the vacuum state on which passive linear optical
operations (described by orthogonal and symplectic
transformations) as well as complete or partial homodyne
measurements can be performed. In principle these might help to
increase the entanglement in $\gamma$, but in the following we
show that this is not the case. We consider the following general
set-up: consider system with CM $\gamma$, ancilla systems in
vacuum state i.e., $\gamma_\mathrm{anc}=\id$, linear passive
interactions (described by a symplectic and orthogonal matrix $O$)
between the system light mode and the ancillas (e.g. beam splitter
between light and ancillary modes), such that the whole system is
described by the CM $\gamma' =
O^T(\gamma\oplus\gamma_\mathrm{anc})O$; clearly,
$\cS(\gamma')=\cS(\gamma)$ and now we show that a Gaussian
measurement does not increase $\cS(\gamma)$: We write $\gamma'$ as
\[
\gamma' = \left( \begin{array}{cc} A'&C'\\ C'^T&B'
\end{array} \right),
\]
where the block matrix $B'$ refers to the ancillary modes to be
measured. Then the resulting state is described by the CM
$\gamma_\mathrm{out}=A'-C'B'^{-1}C'^T$ \cite{GiCi02}. Using the following
characterization of the smallest eigenvalues \cite{HoJo87} it is
straight forward to see that measurement has reduced the squeezing
of the state:
\begin{eqnarray*}
\cS(\gamma_{out}) &=& \mbox{min}_{x\in\CC^n}\left\{
\frac{x^\dagger(A'-C'B'^{-1}C'^T)x}{x^\dagger x} \right\}^{-1}\\
&\leq&\mbox{min}_x\left\{
\frac{x^\dagger(A'-C'B'^{-1}C'^T)x}{x^\dagger(\id+C'B'^{-2}C'^T)
x}
\right\}^{-1}\\
&=&\mbox{min}_x\left\{ \frac{y^\dagger\gamma'y}{y^\dagger y} : y =
{x\choose-B'^{-1}C'^Tx}
\right\}^{-1}\\
&\leq&\mbox{min}_{y\in\CC^{2n}}\left\{
\frac{y^\dagger\gamma'y}{y^\dagger y} \right\} =
\cS(\gamma')\,\,\,\,\hfill\qed
\end{eqnarray*}
Consequently, unsqueezed ancilla systems and Gaussian measurements are
of no help in increasing the squeezing or entanglement in a Gaussian
state.

The preceding discussion does not completely solve the
problem of optimal entanglement generation with a Hamiltonian
$H$, since only one particular initial state (the vacuum) has
been considered.  If, e.g., the initial state of the light field
is squeezed, we have seen in \Sref{SecRate} that better rates can
be achieved (see Fig.~\ref{FRate}), which will translate into larger
entanglement after finite 
times. The methods used above easily yield an upper bound for the
entanglement that can be obtained from initially squeezed states:
Consider an initial product state with squeezing $e^{r_1}$ and
$e^{r_2}$ in systems 1 and 2 and let $r_1\geq r_2$. By the same
arguments as above, after $H$ has acted for a time $t$ the
squeezing in the final state and the negativity are bounded by
$e^{(s_1-s_2)t+r_1}$. We can find a better bound on the achievable
entanglement drawing on results from Ref. \cite{WoEP02}, where it
was shown that the negativity of a two-mode CM $\gamma$ is bounded
by $1/\sqrt{\lambda_1\lambda_2}$, where $\lambda_1,\lambda_2$ are
the two smallest eigenvalues of the $\gamma$. This implies that
\begin{equation}\label{bound}
\cN(\gamma_\mathrm{out})\leq e^{(s_1-s_2)t+(r_1+r_2)/2}, 
\end{equation}
which yields the
dash-dotted curve in Fig.~\ref{FRate}b. This bound is most probably
not tight for $r_k\not=0$, not even as $t\to\infty$.

\begin{figure}
\includegraphics[scale=0.9]{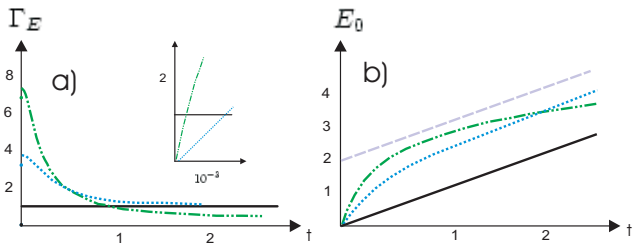}
\caption{\label{Fcounter} (a) The entanglement rate obtained for the
initial state $\gamma_\mathrm{in,2} =
S_{r_1,r_2}\gamma_\mathrm{tms}(t_0/2)S_{r_1,r_2}^T$, where
$S_{r_1,r_2}=\mathrm{diag}(e^{r_1/2},e^{-r_1/2},e^{r_2/2},e^{-r_2/2})$ and
$r_1=r_2=2, t_0=10^{-3}$. The solid line $\Gamma_E=1$ is obtained
with the strategy that optimizes the entanglement rate at each time; 
the dotted line represents the rate obtained for optimal simulation of
$H_\mathrm{tms}$; the ``dot--dot--dashed'' line represents the rate obtained
for the natural Hamiltonian $H_0=X_1X_2$. The inset shows that one has
to ``pay'' with initial 
entanglement rates smaller than the optimal value of 1 to reach a
state that allows for the large rates later on. 
(b) The entanglement created by the different strategies [same styles
for different scenarios as in a)] and the upper bound \Eqref{bound}. }
\end{figure}

One might think that in order to optimize the entanglement after some
finite time $t$ it always suffices to optimize the rate at each time
as for the case of a vacuum input. For qubit systems this was indeed
shown to be true \cite{DuVCLP00}. In contrast, it does not hold for cv systems
as the counterexample depicted in \Fref{Fcounter} shows: We start with
a slightly entangled state with CM $\gamma_\mathrm{in,2}$ which can be
obtained from the two-mode squeezed state $\gamma_\mathrm{tms}(t_0/2)$
squeezing both $X_1$ and $X_2$ by $r_1=r_2$. Then the ``local squeezing
parameter'' $l$ is zero and the optimal rate 
therefore $\Gamma_E=1$. If $t_0$ is small and $r_1, r_2$ large it is
possible to sacrifice some entanglement in order to ``activate'' the
local squeezing thus enhancing the rate later on and obtaining
significantly more entanglement at time $t\gg t_0$.
The difference to the qubit case is related to the fact that in the
cv context not all local transformations are available and hence not
all equally entangled states are locally equivalent. 


\section{Discussion and Conclusion}\label{SecConc}

We have investigated how a quadratic interaction between two
continuous variable systems (as it occurs naturally in
certain quantum optical systems) can be optimally used to perform
several quantum information tasks when certain simple local control
operations (phase space rotations) can be implemented as well.
First we have given necessary and sufficient conditions for the
simulation of a Hamiltonian evolution given a fixed interaction and
fast local rotations. In particular, we have shown that
the naturally occurring Hamiltonian \Eqref{Hint} allows to simulate
all bilinear Hamiltonians and is in fact of the most versatile kind
for this purpose. Moreover we have seen that almost all the
Hamiltonians of the form (\ref{generalHint}) (and in particular
$H_0$) allow to generate all symplectic transformations on two modes,
i.e., the complete group $SP(2,\RR)$ can be generated starting from no
more than the three Hamiltonians $H_0, H_{\mathrm{loc},1},
H_{\mathrm{loc},2}$.

With these results we have addressed the questions of optimal
creation of entanglement and squeezing for a two--mode Gaussian
state using a given interaction of the form (\ref{generalHint}) and local
rotations of the form $H_{\mathrm{loc},i}=g(X_i^2+P_i^2)$, both of
which are available in current experiments. For the case of small
(infinitesimal) interaction times, we have determined the optimal
strategy to increase entanglement or squeezing for any input
state, i.e,  we have derived the maximal entanglement and squeezing rates and
determined the strategies which lead to these maxima. For the
general case (finite interaction time) we have derived the optimal
strategy for the creation of entanglement and squeezing starting
with the vacuum state. We have also shown that (in contrast to qubit
systems) for continuous variables optimizing the entanglement rate is
not necessarily the best way to generate a finite amount of
entanglement.  

There are several interesting applications of our results for quantum
information processing. In particular, we have seen that the beam
splitter Hamiltonian $H_\mathrm{bs} = X_1P_2-P_1X_2$ can be simulated
with an efficiency factor $1/2$ by $H_0$. When acting for a time
$t=\pi$ the Hamiltonian $H_\mathrm{bs}$ generates the swap operation
between the systems 1 and  2, thus performing the ``write-in'' and
``read-out'' operations needed when the atomic ensemble is to be used
as a \emph{quantum memory} for the state of the light mode \cite{KoMP99}.

Another interesting application for atomic ensembles is enabled by
the so-called spin-squeezed states \cite{KiUe93} which have been
prepared experimentally in settings similar to the one described
in this paper \cite{HaSSP99,KuMB00}. It has been shown that these
states allow for a significant increase in the precision of atomic
clocks \cite{SoDCZ00}.  While the methods presented above show
efficient ways to create squeezed atomic states (e.g., by using the
interaction to create squeezing or entanglement optimally and then
project the atoms into a pure squeezed state by measuring the light),
it would also be interesting to find the \emph{optimal} such
procedure. 

Note that the argument in \Sref{SecFinTim} is easily adapted to
similar circumstances. E.g., it was shown in \cite{SoDCZ00} that the
interaction between the atoms of a suitably prepared
Bose-Einstein--condensate (BEC) can be described by the quadratic
Hamiltonian $J_z^2\approx P^2$, which can be used to drive the BEC
into a spin squeezed state.  By the same reasoning as in \Sref{SecFinTim} we
see that after an interaction time $t$ a squeezing of $e^t$ is the
maximum achievable. This shows optimality of the procedure suggested in
\cite{SoDCZ00} (which employs effectively the so-called ``two-axes
counter-twisting'' Hamiltonian).

In summary, we have investigated the capabilities of cv
interaction Hamiltonians $H$. We have shown which other
Hamiltonians can be simulated with such an $H$ and the available control
operations and how to do so efficiently. Then we have derived the
optimal entanglement generation rates achievable with this
Hamiltonian and given an optimal protocol for the generation of
entanglement between the two modes for finite times.

\begin{acknowledgments}
We acknowledge stimulating discussions with Eugene Polzik. This work
was supported in part by the European Union under the project EQUIP
(contract IST-1999-11053).
\end{acknowledgments}

\appendix

\section{Proof of the necessary and sufficient condition for
Hamiltonian simulation}\label{AppSim}

First we prove necessity. If $H$ can simulate $H^{\prime}$
efficiently (\ref{Heff}) has to hold for $\kappa=1$ and
$H_\mathrm{eff}=H^{\prime}$. Therefore and because of (\ref{Kform}) and
(\ref{Transham}) there must exist a probability distribution
$\left\{ p_{i}\right\}  _{i=1}^{n}$ and special orthogonal
matrices $\left\{
R_{i},S_{i}\right\}  _{i=1}^{n}$ such that%
\begin{equation}
\left(
\begin{array}
[c]{cc}%
s_{1}^{\prime} & 0\\
0 & s_{2}^{\prime}%
\end{array}
\right)  =\underset{i=1}{\overset{n}{\sum}}p_{i}R_{i}\left(
\begin{array}
[c]{cc}%
s_{1} & 0\\
0 & s_{2}%
\end{array}
\right)  S_{i}. \label{Matrep}%
\end{equation}
Rotation matrices which should in principle appear on the left
hand side can be removed by left and right multiplication with
corresponding transposed matrices. In (\ref{Matrep}) we assume
these ones to be already included in the $R_{i},S_{i}$ on the
right hand side.

By using the fact that the vector of the diagonal elements of a
product
$R$ $\mathrm{diag}(s_{1},s_{2})$ $S$ can be written as $(R\circ S^{T})(s_{1}%
,s_{2})^{T}$ where $R\circ S^{T}$ denotes the component-wise
(so-called Hadamard) product of matrices we can express the last
equation in compact form
as%
\begin{equation}
\left(
\begin{array}
[c]{c}%
s_{1}^{\prime}\\
s_{2}^{\prime}%
\end{array}
\right)  =\underset{i=1}{\overset{n}{\sum}}p_{i}\left(  R_{i}\circ S_{i}%
^{T}\right)  \left(
\begin{array}
[c]{c}%
s_{1}\\
s_{2}%
\end{array}
\right)  =:N\left(
\begin{array}
[c]{c}%
s_{1}\\
s_{2}%
\end{array}
\right)  . \label{Vecrep}%
\end{equation}
The definition of the matrix $N$ in (\ref{Vecrep}) is obvious.
Using that all matrices $R_{i},S_{i}$ are elements of $SO(2,\RR)$
it can be seen
easily that%
\begin{align*}
N_{11}=N_{22},\quad N_{12}  &  =N_{21}\quad\text{and}\\
\left|  N_{11}\pm N_{21}\right|   &  \leq1.
\end{align*}

Conditions (\ref{Cond}) follow now directly from (\ref{Vecrep})
and these
properties of $N$:%
\[
s_{1}^{\prime}+s_{2}^{\prime}=(N_{11}+N_{21})(s_{1}+s_{2})\leq s_{1}+s_{2}%
\]
The same holds identically for all plus signs replaced by minus
signs proving necessity.

To demonstrate sufficiency we show that conditions (\ref{Cond})
guarantee the existence of a matrix $N$ as in (\ref{Vecrep}) which
in turn admits to connect the primed and unprimed restricted
singular values as in (\ref{Matrep}). This provides an efficient
simulation protocol of the form (\ref{Simprot}).

Given $s_{1},s_{2}$ and $s_{1}^{\prime},s_{2}^{\prime}$ fulfilling
(\ref{Cond}) we can for the time being assume that $s_{1}\neq
|s_{2}|$ and
define%
\begin{align*}
N  &  :=\left(
\begin{array}
[c]{cc}%
e & f\\
f & e
\end{array}
\right),\,\, \mbox{where} \\
e  &  =\frac{s_{1}s_{1}^{\prime}-s_{2}s_{2}^{\prime}}{s_{1}^{2}-s_{2}^{2}%
},\quad f=\frac{s_{1}s_{2}^{\prime}-s_{2}s_{1}^{\prime}}{s_{1}^{2}-s_{2}^{2}%
}.
\end{align*}
With this definition we have $(s_{1}^{\prime},s_{2}^{\prime})^{T}%
=N(s_{1},s_{2})^{T}$. Next we have to show that $N$ can be written
as a convex sum of Hadamard products of rotation matrices which is
in fact exactly what inequalities (\ref{Cond}) ensure.

It is again easy to check that if $\left|  e\right|  +\left|
f\right|  \leq1$ we can find probabilities $\left\{
p_{i}:p_{i}\geq 0,\sum_{i=1}^{4}p_{i}\right\}  _{i=1}^{4}$ such
that $e=p_{1}-p_{2}$ and
$f=p_{3}-p_{4}$ and therefore%

\begin{equation}
\begin{array} [c]{c}
N=p_{1}\left(\begin{array}
[c]{cc} 1 & 0\\
0 & 1 \end{array} \right) \circ\left(\begin{array} [c]{cc}
1 & 0\\
0 & 1
\end{array}
\right)  +p_{2}\left(\begin{array} [c]{cc}
1 & 0\\
0 & 1
\end{array}
\right)  \circ\left(\begin{array} [c]{cc}
-1 & 0\\
0 & -1
\end{array}
\right)  \\
+p_{3}\left(\begin{array} [c]{cc}
0 & 1\\
-1 & 0
\end{array}
\right)  \circ\left(\begin{array} [c]{cc}
0 & 1\\
-1 & 0
\end{array}
\right)  +p_{4}\left(\begin{array} [c]{cc}
0 & 1\\
-1 & 0
\end{array}
\right)  \circ\left(\begin{array} [c]{cc}
0 & -1\\
1 & 0
\end{array}\right)  .
\end{array}
\label{Dec}
\end{equation}
This decomposition of $N$ allows to pass from (\ref{Vecrep}) to
(\ref{Matrep}) conserving the diagonal structure as can be checked
easily. Thus it suffices to show how (\ref{Cond}) implies $\left|
e\right|  +\left|  f\right|  \leq1$. Multiplying the first
[second] line of (\ref{Cond}) by $(s_{1}-s_{2})$
$[(s_{1}+s_{2})]$ yields respectively%
\begin{align*}
s_{1}^{2}-s_{2}^{2} &  \geq(s_{1}s_{1}^{\prime}-s_{2}s_{2}^{\prime}%
)+(s_{1}s_{2}^{\prime}-s_{2}s_{1}^{\prime}),\\
s_{1}^{2}-s_{2}^{2} &  \geq(s_{1}s_{1}^{\prime}-s_{2}s_{2}^{\prime}%
)-(s_{1}s_{2}^{\prime}-s_{2}s_{1}^{\prime}).
\end{align*}
The first term on the right hand sides is nonnegative due to
premise
(\ref{Resvals}) such that these inequalities are equivalent to%
\[
s_{1}^{2}-s_{2}^{2}\geq\left|
s_{1}s_{1}^{\prime}-s_{2}s_{2}^{\prime}\right| +\left|
s_{1}s_{2}^{\prime}-s_{2}s_{1}^{\prime}\right|
\]
which is, regarding the definition of $e$ and $f$, exactly what we
had to show and proves sufficiency for the case $s_{1}\neq
|s_{2}|$.

The complementary cases $s_{1}=|s_{2}|$ turn out to be trivial,
since conditions (\ref{Cond}) then require
$s_{1}^{\prime}=s_{2}^{\prime}=s_{1}$ or
$s_{1}^{\prime}=-s_{2}^{\prime}=s_{1}$ respectively and this means
that we can exclusively simulate Hamiltonians where
$H^{\prime}=$.$\left( U\otimes V\right)  H\left(
U^{\dagger}\otimes V^{\dagger}\right) $ for some local rotations
$U\otimes V$, i.e. $H^{\prime}$ has to be - in this sense -
locally equivalent to $H$. Hence, nothing has to be shown in this
case. $\square$

We point out that this proof provides the possibility to construct
simulation protocols explicitly. Given $H$ and $H^{\prime}$ one
has to calculate the decomposition in \ref{Dec}. Then the
probabilities and rotations appearing there will fix the time
steps $t_{i}$ and control operations $U_{i}\otimes V_{i}$ in
(\ref{Simprot}). As can be seen such a protocol will contain at
most three intervals of interaction and control operations being
rotations about
$\pm\pi/2$ and $\pi$.

\section{Gate Simulation}\label{AppGSim}

To show that any unitary $U=\exp(-iG)$ where $G$ is quadratic expression in
the operators $X_{k},P_{k}$ can be decomposed as given in (\ref{Decomp}) we
will proceed in three steps:

(i) As shown in \cite{SiMD94,Brau99} any such $U$ can be
decomposed into a sequence of one passive transformation, single
mode squeezing and another passive transformation. That is to say
the symplectic matrix $S$ corresponding to the unitary
transformation $U$ can be decomposed as $S=OD\widetilde{O}$ where
$O,\widetilde{O}$ are orthogonal, symplectic and, therefore,
passive transformations and the diagonal matrix
$D=\mathrm{diag}(e^{\alpha+\beta },e^{-\left(  \alpha+\beta\right)
},e^{\alpha-\beta},e^{-\left(  \alpha -\beta\right)  })$ amounts
to local squeezing. Note that this is basically a singular value
decomposition of $S$.

(ii) Passive transformations contain essentially beam-splitter
transformations
and local rotations and it is well known from quantum optics that any such
transformation on two modes can be decomposed into a sequence of a pair of local
rotations,
one beam-splitter operation and another pair of local rotations. Thus, a
unitary $U_{O}$ corresponding to a orthogonal symplectic transformation $O$
can be decomposed as $U_{O}=(V\otimes W)U_{\mathrm{bs}}(t_{0})(\widetilde
{V}\otimes\widetilde{W})$ where $U_{\mathrm{bs}}(t)$ is defined in
(\ref{Bs}).

(iii) What is left to show is how to attain single mode squeezing. For
this we split the matrix $D$ into two components, $D=\mathrm{diag}(e^{\alpha
},e^{-\alpha},e^{\alpha},e^{-\alpha})\mathrm{diag}(e^{\beta},e^{-\beta
},e^{-\beta},e^{\beta})$ and show how each of them can be attained by
means of
beam-splitters and two-mode squeezing. Let us denote by $\overline{U}_{\mathrm{bs}%
}(t)$ and $\overline{U}_{\mathrm{tms}}(t)$ the variants of beam splitter and
two-mode squeezing operators which are attained from (\ref{Bs}) and
(\ref{Tms}) respectively by locally rotating $X_{2}\rightarrow P_{2}%
,P_{2}\rightarrow-X_{2}$. Then it can be easily shown that the
sequence
$\overline{U}_{\mathrm{bs}}(-\pi/4)U_{\mathrm{tms}}(\alpha)\overline
{U}_{\mathrm{bs}}(\pi/4)$ generates a symplectic transformation
$\mathrm{diag}%
(e^{\alpha},e^{-\alpha},e^{\alpha},e^{-\alpha})$ and $U_{\mathrm{bs}}%
(-\pi/4)\overline{U}_{\mathrm{tms}}(\beta)U_{\mathrm{bs}}(\pi/4)$
correspondingly
$\mathrm{diag}(e^{\beta},e^{-\beta},e^{-\beta},e^{\beta})$.

Collecting things together and ordering all passive components as
in (ii), i.e. such that it contains only one application of a
beam-splitter operation, decomposition (\ref{Decomp}) follows
immediately.


\end{document}